\documentclass{egpubl}
\usepackage{sca2020}

\usepackage[T1]{fontenc}
\usepackage{dfadobe}  

\biberVersion
\BibtexOrBiblatex
\usepackage[backend=biber,bibstyle=EG,citestyle=alphabetic,backref=true]{biblatex} 
\electronicVersion
\PrintedOrElectronic
\ifpdf \usepackage[pdftex]{graphicx} \pdfcompresslevel=9
\else \usepackage[dvips]{graphicx} \fi

\usepackage{booktabs} 
\usepackage{mathtools}
\usepackage{subfig}
\usepackage{lipsum}
\usepackage{amssymb}
\usepackage[ruled]{algorithm2e} 

\SetAlFnt{\small}
\SetAlCapFnt{\small}
\SetAlCapNameFnt{\small}
\SetAlCapHSkip{0pt}

\title[Optimized Processing of Localized Collisions in Projective Dynamics]%
      {Optimized Processing of Localized Collisions in Projective Dynamics}

\author[Q. Wang et al.]
{\parbox{\textwidth}{\centering Qisi Wang\textsuperscript{1}, Yutian Tao\textsuperscript{1}, Eric Brandt\textsuperscript{1}, Court Cutting\textsuperscript{2}, and Eftychios Sifakis\textsuperscript{1,3} \\ \bigskip
\textsuperscript{1} University of Wisconsin, Madison \hspace*{.1in}
\textsuperscript{2} New York University Medical Center \hspace*{.1in}
\textsuperscript{3} Weta Digital}}

\begin{document}

\newcommand\todo[1]{\textcolor{red}{TODO: #1}}

\newcommand*{\tran}{^{\mkern-1.5mu\mathsf{T}}}
\newcommand*{\invtran}{^{\mkern-1.5mu\mathsf{-T}}}
\newcommand*{\invse}{^{\mkern-1.5mu\mathsf{-1}}}

\newcommand\scalar[1]{#1}
\newcommand\aggvec[1]{\vec{\mathbf{#1}}}
\newcommand\aggsca[1]{\mathbf{#1}}
\newcommand\aggmat[1]{\mathbf{#1}}
\newcommand\mat[1]{\mathrm{#1}}
\newcommand\vect[1]{\vec{#1}}
\newcommand\elemset[1]{\mathcal{#1}}

\newcommand\iter[1]{_{(#1)}}
\newcommand\indx[1]{^{(#1)}}

\teaser{
 \centering{
 \includegraphics[height=1.49in]{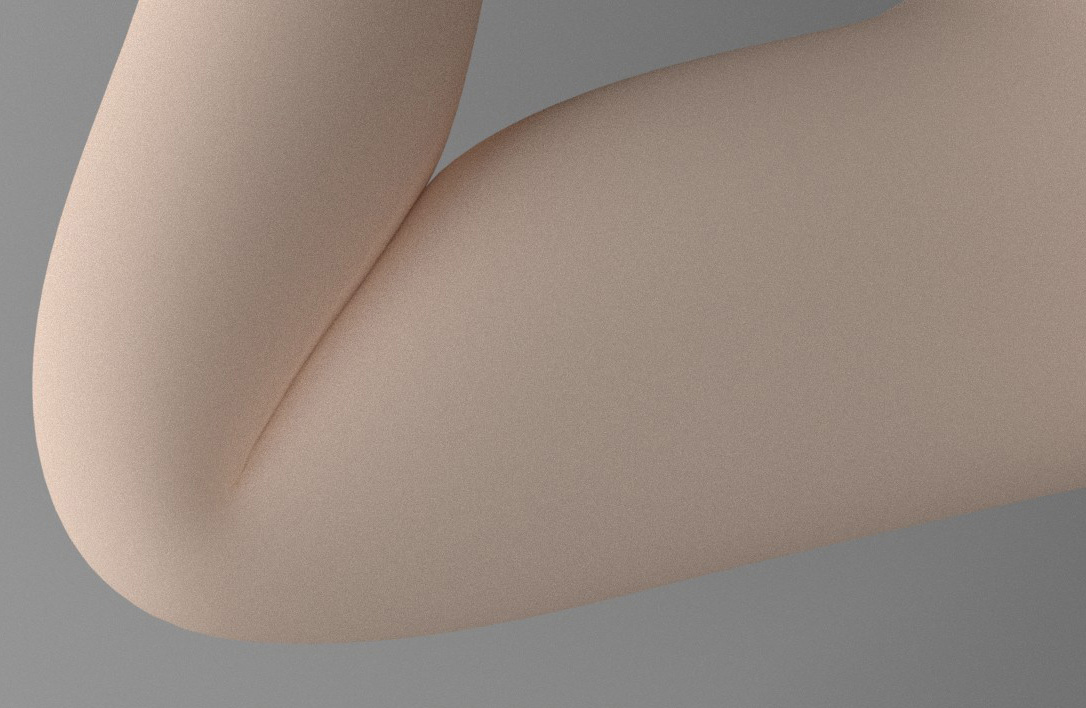}
  \includegraphics[height=1.49in]{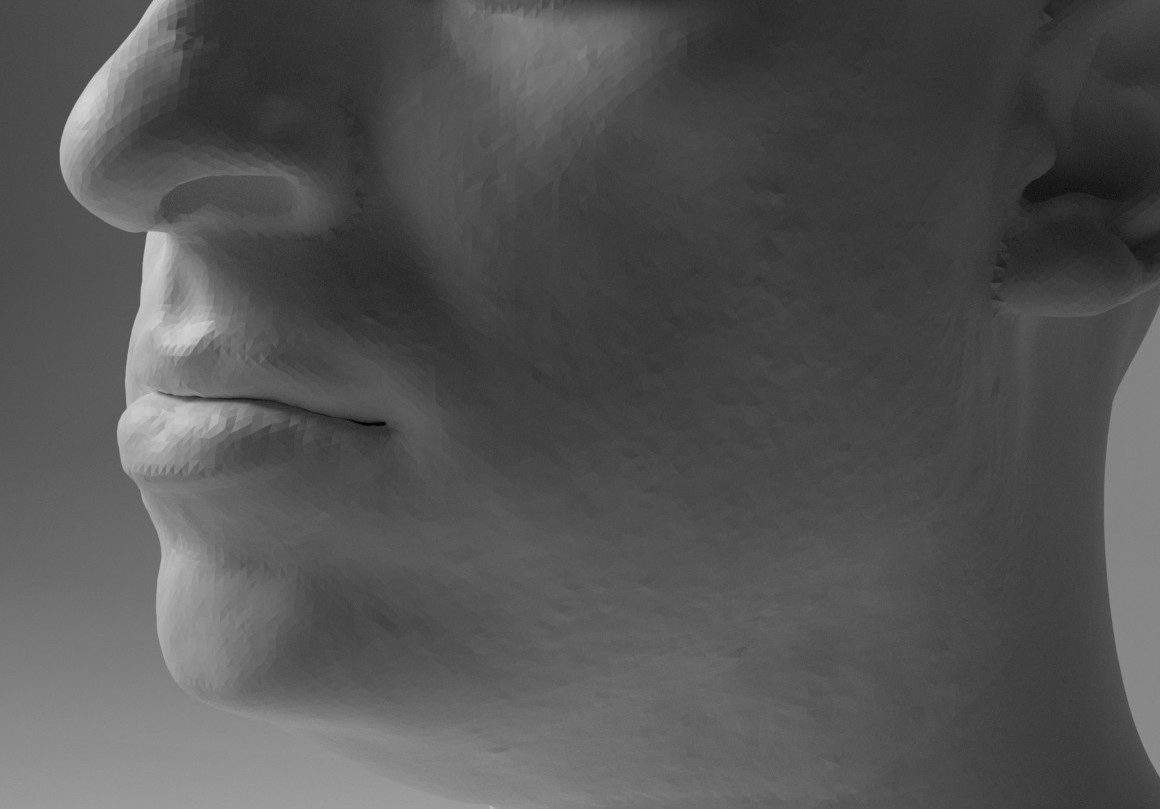}
  \includegraphics[height=1.49in]{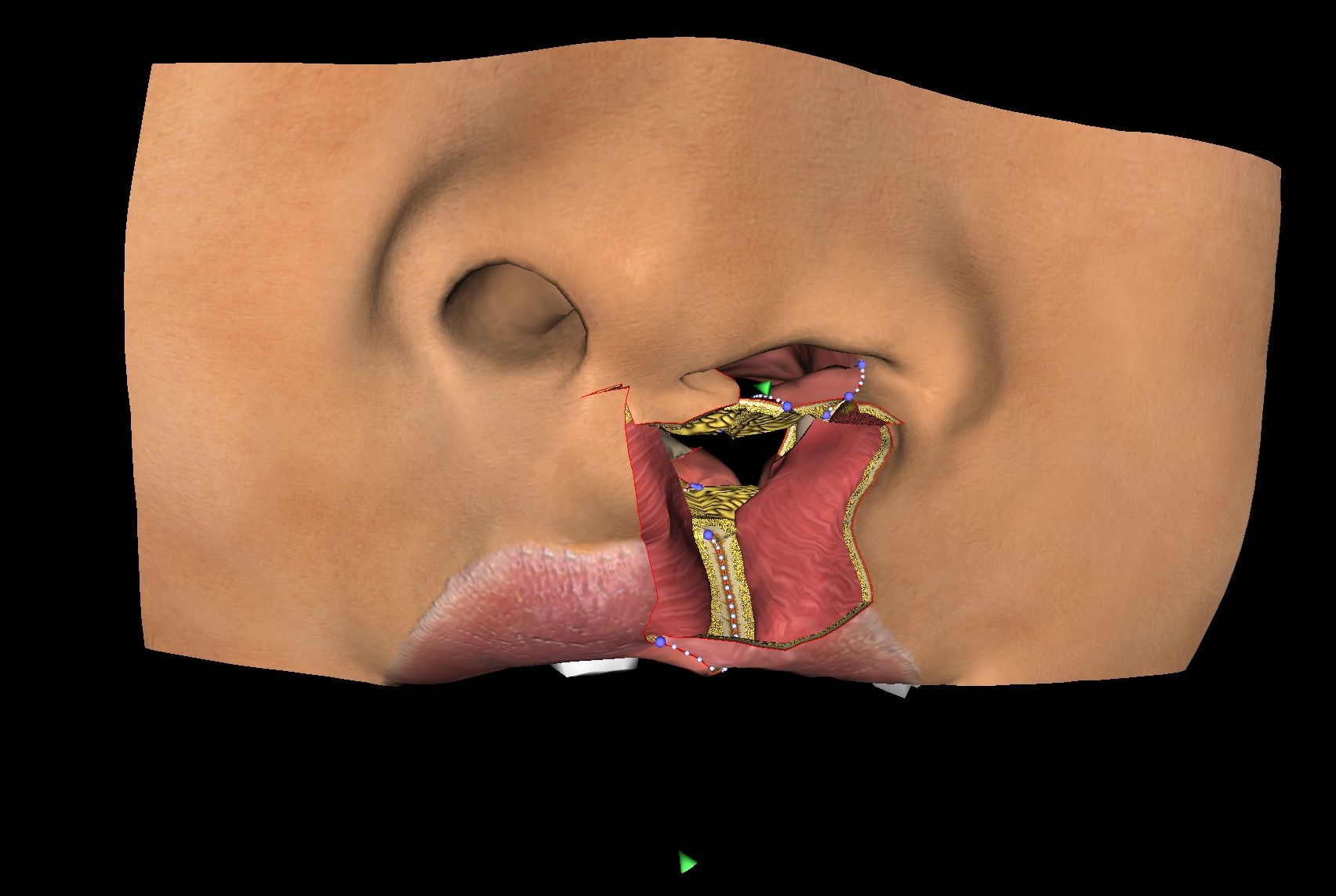}
}
\caption{Demonstrations of collision handling in our framework. Left: An elbow model, embedded in a tetrahedral simulation mesh with 596K elements. Middle: A face model (644K tetrahedral elements) brought into a self-colliding configuration by articulating the mandible. Right: Simulation of a cleft lip and palate repair in a virtual surgery simulator (468K tetrahedral elements). Persistent collision occurs between the lip and the gum/teeth in the maxilla. Simulation rates for all these examples range between $3$-$8$fps, with full collision handling.} 
  \label{fig:teaser}
}

\maketitle
\begin{abstract}
We present a method for the efficient processing of contact and collision in volumetric elastic models simulated using the Projective Dynamics paradigm. Our approach enables interactive simulation of tetrahedral meshes with more than half a million elements, provided that the model satisfies two fundamental properties: the region of the model's surface that is susceptible to collision events needs to be known in advance, and the simulation degrees of freedom associated with that surface region should be limited to a small fraction (e.g. 5\%) of the total simulation nodes. Despite this conscious delineation of scope, our hypotheses hold true for common animation subjects, such as simulated models of the human face and parts of the body. In such scenarios, a partial Cholesky factorization can abstract away the behavior of the collision-safe subset of the face into the Schur Complement matrix with respect to the collision-prone region. We demonstrate how fast and accurate updates of penalty-based collision terms can be incorporated into this representation, and solved with high efficiency on the GPU. We also demonstrate the opportunity to iterate a partial update of the element rotations, akin to a selective application of the local step, specifically on the smaller collision-prone region without explicitly paying the cost associated with the rest of the simulation mesh. We demonstrate efficient and robust interactive simulation in detailed models from animation and medical applications. 

\ccsdesc[500]{Computing methodologies~Physical simulation}
\ccsdesc[500]{Computing methodologies~Collision detection}


\printccsdesc   
\end{abstract}  

\section{Introduction}

Projective Dynamics \cite{bouaziz2014projective} is a popular, robust, and efficient iterative scheme for interactive simulation of models govered by the corotational elasticity constitutive model. Equivalent in principle to a quasi-Newton scheme \cite{Liu2017}, Projective Dynamics (PD) often delivers significant advantages against traditional Newton-style implicit schemes, in terms of stability and efficiency. Robust and stable simulation is guaranteed by casting each time step of PD as an optimization problem, in which both of its alternating components (e.g. the ``local'' and ``global'' step) is assured to decrease monotonically. Such guarantees do not exist in a traditional Newton scheme in the absence of linesearch failsafes.
Efficiency in Projective Dynamics largely stems from the fact that the modified Hessian it uses when viewed as a quasi-Newton scheme is a constant Laplacian-like matrix that can be prefactorized and efficiently solved using forward/backward substitution. This is in contrast to the true Hessian of full-Newton schemes which varies with deformation and can also become indefinite, limiting the available options for high-performance, yet robust solvers. 



\paragraph*{Challenges} While Projective Dynamics enjoys these benefits, it is not without limitations and challenges. Some of these are associated with its specific affinity to corotated elasticity (or close variants \cite{Ichim2017}) making it less than ideal to pair with generic material models. But more importantly, the presence of collisions has the tendency to clash with many of the preconditions that contribute to the robustness, efficiency, and favorable convergence of Projective Dynamics.
Collision resolution for volumetric elastic objects, especially in the context of implicit or quasistatic simulations, is most frequently handled via the imposition of penalty forces on parts of the mesh that penetrate into prohibited regions \cite{teschner2005collision,Mcadams2011,mitchell2015non}. This response is materialized in the form of short-lived zero-restlength springs that connect points on the model surface found to be colliding with the closest surface point on the ``other side'' of the collision. As such, the proper treatment of such spring forces would be to incorporate them into the Laplacian-like matrix in the global step of Projective Dynamics. This can be absolutely detrimental to the ability of Projective Dynamics to use a factorization-based direct solver, since the update cost (say, of a Cholesky decomposition) would be prohibitive in an interactive application. Models with hundreds of thousands of elements, which could otherwise be simulated using direct solvers for the global step in interactive rates, would no longer enjoy such performance if the pre-factorization opportunity is compromised.

Preserving the ability to use a direct method for the global step typically comes with some type of compromise. It is possible, for example, to build the matrix of the global step under the premise that \emph{all} collision sites used in detection (often referred to in the literature as ``collision proxies'' \cite{Mcadams2011}) are engaged in active collision, while the right-hand side can be built with only active proxies taken into consideration; this option was discussed by the original proposers of Projective Dynamics \cite{bouaziz2014projective}. Although this approach retains stability, it adds unnecessary drag on collision proxies that are not actually colliding, and is problematic for self-collisions when the pattern of interaction between colliding parts of the mesh cannot be statically inferred. Later work \cite{Ichim2016} proposed adding linear equality constraints associated with active collisions to the minimization problem in the global step of PD, and using a Schur complement with respect to the constraint equations to build a smaller dense system, with the dimension of the active constraints. Although this approach is quite flexible, it requires a somewhat expensive update of the Schur complement at each iteration, and is only practical for a relatively small number (at most a few hundred) active collision proxies. Our approach also leverages Schur complements, but in a very different context as we will see.

\paragraph*{Proposed method and Scope} In this paper, we propose a new and distinctive approach to reconciling collision processing with the philosophy of Projective Dynamics. Our method safeguards the strong robustness guarantees of PD and its ability to use an accurate, direct solver for the global step,while retaining very attractive performance on models of substantial resolution, but there is a price we consciously have to accept: We commit to an upfront narrowing of our scope of applicability to simulation scenarios that satisfy the following two conditions: (1) We must know in advance which sections of the object's surface are likely, by-and-large, to ever be engaged in collision. We shall call this the \emph{collision-prone} region; (2) The simulation nodes that are associated with collision proxies (either by \emph{being} collision proxies themselves, or \emph{embedding} them) in the collision-prone region should only be a small fraction of the total nodes in the simulation mesh, e.g. ideally less than $5\%$ of a volumetric mesh with more than 100K vertices as in our examples.

It is not difficult to identify simulation scenarios that satisfy these stipulations -- and others that would not. Figure \ref{fig:teaser} illustrates such scenarios featured in our demonstrations. Models of the human face would be a prime candidate, if we accept the modeling hypothesis that collisions will only be handled on the immediate vicinity of the mouth. For reasonably resolved face meshes with several hundred of thousand tetrahedral elements, it is easy to localize the collision-prone region to no more than a few thousand nodes. On the other hand, this assumption would not hold if we intended to collide the face with external objects without restricting where the contact takes place. Body models would also satisfy this stipulation if we only targeted collisions that appear around joints: the elbow, the underarm area, the region behind the knee, etc. Again, considering collisions with external objects, or non-local self-collisions (e.g. hand touching the torso) would break our hypothesis. 

If, however, these modeling assumptions do hold true for our simulation task, we are presented with a very clear opportunity for highly-optimized processing and accurate treatment of collisions within Projective Dynamics, while retaining the stability and convergence of direct solvers. Our method can then separate our simulation mesh in collision-safe, and collision-prone regions, and use a partial Cholesky factorization to reduce the computation that needs to occur during the global step into a problem that \emph{only involves the collision-prone degrees of freedom}. This localized problem is a linear system of equations, using the Schur Complement of the traditional global step Laplacian (with respect to the collision-prone nodes) as its coefficient matrix; the core benefit is that updates to the overall scheme due to activation or deactivation of collision proxies is purely \emph{sparse, additive updates} to the Schur Complement. 
Our formulation also affords the opportunity to update the optimal rotations of elements in the collision-prone region at the same time that we repeat collision detection, but without explicitly updating the collision-safe region and at drastically reduced cost. For models with a resolution in the order of half a million tetrahedral elements we can perform accurate penalty-based collision handling at no more than twice (and often much less) the cost of the same model simulated without collisions. 

Finally, in delineating our scope, we clarify that our method presumes that using a direct solver as opposed to an iterative scheme for the global step is something the user seeks to preserve. This is often motivated by the accuracy and robustness of a direct solver, and avoiding the need to fine-tune the iterative scheme to the model resolution, stiffness of constraints, or abrupt nature of motion.
We should disclose, however, that in our experience for models with significantly lower resolution than what we target (e.g. in the order of 50K-100K elements) or in dynamic simulation aided by inertia, we have found the convergence of iterative methods to be very adequate even with modest iteration count. 
In such instances, an accelerated iterative solver \cite{Komaritzan2019} could be best suited to solving the global step. In section \ref{sec:results} we comment further on benefits of direct solvers for higher resolution models, such as the ones we  target.

\section {Related Work}

\paragraph*{Corotated elasticity} Simulation of deformable bodies using corotated elasticity strikes a good balance between respecting nonlinearity and rotational invariance, while revealing opportunities for interactive simulation. The principle of Corotated Elasticity first materialized in warped stiffness methods \cite{muller2002stable}, and later made rotationally invariant \cite{muller2004interactive}, and robust to inversion \cite{irving2004invertible} and indefiniteness of the stiffness matrix \cite{teran2005robust}. Analytic second derivatives of the corotated energy allowed improved convergence of Newton Methods \cite{Mcadams2011,chao2010simple} while the derivative singularity of the model around highly compressed configurations was treated with appropriate modifications \cite{stomakhin2012energetically}. 

\paragraph*{Projective Dynamics} Targeting corotated elasticity as a material model, the concept of Projective Dynamics \cite{bouaziz2014projective} has enjoyed significant adoption and evolution. Analyzed as a quasi-Newton scheme \cite{Liu2017} and related to ADMM optimization \cite{Narain2016}, it has been used for developing damping models \cite{Jaillet2018}, elastic rod simulations \cite{Soler2018}, face animation \cite{Ichim2017}, motion control using volumetric actuators \cite{Lee2018}, skinning simulation \cite{Komaritzan2018, Komaritzan2019} and reduced models \cite{Brandt2018}. The relation between Projective Dynamics and ADMM has also been investigated \cite{Narain2016,overby2017admm}, allowing more general constitutive models and constraints to be used, with iterative solvers utilized for the global step, albeit typically demonstrated at more modest resolutions than we use. Chebyshev iteration has also been used to tackle the global step \cite{wang2015chebyshev}, allowing efficient GPU implementation, albeit carrying weaker guarantees for robustness relative to direct solvers. Among these approaches, we find that iterative methods based on GPU-acclerated Conjugate Gradients solvers \cite{Komaritzan2018, Komaritzan2019} are the closest in scope to our work; as we discuss in Section \ref{sec:results} such schemes would be preferable for models of more modest resolution than ours (we use about half a million elements), where CG would converge well and not require localization of collisions.

\paragraph*{Skinning and collisions} Collision processing for volumetric objects can leverage more flexible, and occasionally more performant techniques than those used for cloth simulation, due to its ability to recover from tangled configurations. Detection responses leveraging implicit geometry representations have seen significant adoption \cite{teschner2005collision, Mcadams2011, mitchell2015non}, and typically employ penalty force formulations for collision response. Recent skinning methods that focus on interactive simulation include implicit skinning \cite{vaillant2013implicit}, Delta mush \cite{Le2019}, methods that exploit the Projective Dynamics concept \cite{Komaritzan2018, Komaritzan2019}, Position-Based Dynamics \cite{abu2015position}, and subspace deformation \cite{teng2014simulating}, often in conjunction with Projective Dynamics \cite{lan2020medial}. Contact and collision for muscle-based skinning simulations have also leveraged volume-preserving fiber primitives \cite{angles2019viper} and simplified yet anatomy-inspired muscle primitives coupled with the Implicit Skinning concept \cite{roussellet2018dynamic}. Finally, using Projective Dynamics in simulations involving frictional contact \cite{ly2020projective} was recently explored.

\section{Technical Background}

\subsection{Notation}

In this paper, we denote variables that represent aggregate quantities (e.g. concatenated lists of a physical property on \emph{all nodes}, or \emph{all elements}) by using boldface type. These aggregate quantities can be either scalar or vector, which are differentiated by an arrow over the variable for vector quantities. For example $\aggsca{y}$ might be the $y$-coordinates of all vertices in a mesh, while $\aggvec{v}$ might be all 3d vertices of the mesh, consisting of the three components $\mathbf{v^{(1)}}, \mathbf{v^{(2)}}, \mathbf{v^{(3)}}$. Subscripts in parenthesis denote iteration numbers, for example $\aggvec{x}\iter{2}$ might be the 3D values of all the mesh vertices after the second iteration of an algorithm. Matrices are capital Roman letters (non-bolded). Aggregate matrices are boldfaced versions of the single matrix notation. 

\subsection{Projective Dynamics}

We start by reviewing the mathematical formulation for Projective Dynamics \cite{bouaziz2014projective}, with the slight modification that we attempt to cast the description slightly more in the language of continuum mechanics (including concepts such as stress and force), instead of the style used by the original authors, which was more attuned to a Computational Geometry and Optimization viewpoint.

When simulating an elastic body using the Finite Element Method, the body is first discretized with a volumetric mesh composed of many discrete elements (tetrahedra in our case). Assuming linear tetrahedral elements \cite{sifakis2012fem}, we can compute a deformation gradient, $F_e(\aggvec{x})$, which is constant in each element $e$ and is a linear function of the deformed locations $\aggvec{x}$ of the mesh vertices. The constitutive model of corotated elasticity defines the energy density $\Psi(F)$ as a function of the deformation gradient:
\begin{equation}\label{eq:psi_f}
\Psi(\mat{F}) = \mu || \mat{F} - \mat{R}(\mat{F}) ||_F^2 + \frac{\lambda}{2}\text{tr}^2(\mat{S}(\mat{F})-\mat{I})
\end{equation}
where $\mat{R}(\mat{F}), \mat{S}(\mat{F})$ are the rotational and symmetric components of the deformation gradient given by the polar decomposition $\mat{F} = \mat{R}\mat{S}$, and $\mu, \lambda$ are the Lam\'e coefficients. In keeping with the typical mode of use of Projective Dynamics, we  omit the $\lambda$ term by setting this value to zero.
We draw attention to the important detail that $\mat{R}$ is a dependent function of $\mat{F}$  in this formulation.
Projective Dynamics suggests an alternative formulation of Equation (\ref{eq:psi_f}) where $\mat{R}$ is no longer a function of $\mat{F}$, but rather an independent variable:
\begin{equation}\label{eq:psihat_f}
\hat{\Psi}(\mat{F},\mat{R}) = \mu || \mat{F} - \mat{R} ||_F^2
\end{equation}

The fundamental observation at the core of Projective Dynamics is that the conventional description of the constitutive model's force density function, $\Psi(\mat{F})$, is equal to the minimum over all rotation matrices $\mat{R}$ of the Projective Dynamics energy density  $\hat{\Psi}(\mat{F},\mat{R})$:
$$
\Psi(\mat{F}) = \min_{\mat{R} \in \text{SO}(3)} \hat{\Psi}(\mat{F},\mat{R})
$$

We transition from the (constant) energy density function across each element to an integrated energy function for the same element by  multiplying by the (undeformed) volume of each element:
$$
E_e(\mat{F}_e) = \text{Vol}_e \Psi(\mat{F}_e) = \min_{\mat{R}_e \in \text{SO}(3)}\text{Vol}_e\hat{\Psi}(\mat{F}_e, \mat{R}_e)
=\min_{\mat{R}_e \in \text{SO}(3)}\hat{E}_e(\mat{F}_e, \mat{R}_e)
$$
where we have defined $\hat{E}_e(\mat{F}, \mat{R}) := \text{Vol}_e\hat{\Psi}(\mat{F}, \mat{R})$.
The overall energy of the entire body (with rotations momentarily regarded as independent variables) is the sum of all elemental energies:
\begin{equation}
\hat{E}(\aggvec{x}, \aggmat{R}) = \sum_e \hat{E}_e(\mat{F}_e(\aggvec{x}), \mat{R}_e)
\label{eqn:energyHat}
\end{equation}
from which we can recover the conventional discrete corotated energy for the entire mesh by minimizing rotations over all elements:
$$
E(\aggvec{x}) = \min_{\aggmat{R}}\hat{E}(\aggvec{x}, \aggmat{R})
$$
where it is implied (for brevity of notation) that the minimum is taken over an aggregate $\aggmat{R}$ of matrices that are all rotations (i.e. in SO(3)). We may intuitively interpret the energy $\hat{E}(\aggvec{x}, \aggmat{R})$ as a separate consitutive model from corotational elasticity, where each element's matrix $R_e$ is no longer functionally tied to its deformation gradient, but is simply an element-specific simulation parameter.

Projective Dynamics is usable both in a quasistatic, as well as an implicit Backward Euler time integration scheme, as both cases are ultimately cast in very similar optimization problems. For simplicity of exposition, in this paper we focus on the quasistatic case, with the understanding that our methodology remains fully applicable in the case where Backward Euler is used. In a quasistatic simulation, the deformable system evolves as to satisfy a force equilibrium condition, or equivalently in pursuit of a minimizer for the energy:

\begin{equation}\label{eq:min_e}
\min_{\aggvec{x}} E(\aggvec{x}) = \min_{\aggvec{x}, \aggmat{R}} \hat{E}(\aggvec{x}, \aggmat{R})
\end{equation}

Therefore, the quasistatic evolution can be seen as a minimization of the modified energy $\hat{E}$ jointly over both \emph{independent} parameters $\aggvec{x}$ and $\aggmat{R}$. Projective Dynamics chooses to conduct this minimization by alternating the following two steps until convergence:

\begin{description}
\item[Local Step] Treat $\aggvec{x}$ as constant, and minimize $\hat{E}$ over all $\aggmat{R}$. 
\item[Global Step] Treat $\aggmat{R}$ as constant and minimize $\hat{E}$ over all $\aggvec{x}$.
\end{description}

The stability property of PD results from the fact that each of the aforementioned minimization steps can be iterated while guaranteeing that the energy will monotonically decrease after the application of each one. The local step of minimizing $\aggmat{R}$ is actually multiple independent steps of minimizing $\mat{R}_e$ separately for \textit{each} element. Because these $\mat{R}_e$ are independent, the problem is highly parallel. The solution to the minimization of $\mat{R}_e$ of each element is obtained via the        Orthogonal Procrustes Problem and yields the minimizer $\mat{R}_e = \mat{U}_e(\mat{V}_e)\tran$ where $\mat{F}_e = \mat{U}_e \Sigma_e (\mat{V}_e)^T$ is the SVD of $\mat{F}_e$.
The minimization associated with the \emph{global step} will be handled by an application of a Newton-Raphson procedure, producing an iterative update sequence
$
\aggvec{x}\iter{k+1} \leftarrow \aggvec{x}\iter{k} + \delta \aggvec{x}
$
where $\delta x$ is computed by solving:
\begin{equation} \label{eq:nr}
\frac{\partial^2\hat{E}}{\partial \aggvec{x}^2}\bigg\rvert_{\aggvec{x}_{(k)}}\delta \aggvec{x} = -\frac{\partial\hat{E}}{\partial \aggvec{x}}\bigg\rvert_{\aggvec{x}_{(k)}}
\end{equation}

Let us examine the components of (\ref{eq:nr}) more closely. On the left hand side, we recognize the second derivative of the reformulated energy from Equation (\ref{eqn:energyHat}). Having treated the rotations $\aggmat{R}$ as an independent parameter, this energy is a pure quadratic function of positions $\aggvec{x}$, thus the Hessian is a constant matrix.
When the expression in Equation (\ref{eqn:energyHat}) is interpreted as a modified constitutive model with $\aggmat{R}$ being an independent parameter, this Hessian would be intuitively associated with the ``stiffness matrix'' of this material model, which we denote as $\mat{K}_{\text{el}}$ (with the subscript denoting that this is the ``elastic'' energy, contrasted to collision-spawned contributions discussed later). Similarly on the right-hand side, we recognize the term $-\frac{\partial\hat{E}}{\partial \aggvec{x}}\bigg\rvert_{\aggvec{x}_{(k)}}$ as $\aggvec{f}_{\text{el}}(\aggvec{x}_{(k)})$, which are the aggregate elastic forces computed on the mesh nodes from this constitutive model, at position $\vec{x}_{(k)}$ \cite{Sifakis2015}. This allows us to write an equivalent expression:
\begin{equation} \label{eq:k_el}
\mat{K}_{\text{el}} \delta \aggvec{x} = \aggvec{f}_{\text{el}}(\aggvec{x}_{(k)})
\end{equation}

The stiffness matrix $\mat{K}_{\text{el}}$ can be computed either in the fashion of the Projective Dynamics formulation \cite{bouaziz2014projective}, or by following the Finite Element route which would produce exactly the same result. For the force $\aggvec{f}_{\text{el}}(\aggvec{x}_{(k)})$ however, we opt for a computation using the Finite Element paradigm: On each particular element, $e$, we start by calculating the deformation gradient, $\mat{F}_e$, then calculating the first Piola stress tensor, $P(\mat{F}_e)$, which in turn is used to calculate the force, $\aggvec{f}_e$ \cite{Sifakis2015}. The first Piola stress tensor will be given by $P = \frac{\partial\hat{E}}{\partial F} = 2\mu (\mat{F} - \mat{R})$ which is seemingly the same as that of corotated elasticity \cite{Mcadams2011}, with the caveat that $\mat{R}$ is still treated as an independent parameter rather than a function of $\mat{F}$ (the two will have been brought in sync, by virtue of the preceding local step). 

We conclude our review of the core Projective Dynamics theory with some implementation-minded observations. The stiffness matrix $\mat{K}_{\text{el}}$ has been shown \cite{bouaziz2014projective} to be block diagonal (in the sense that it has no cross-terms that straddle different coordinate components among $\aggsca{x}, \aggsca{y}$, and $\aggsca{z}$) and also all three diagonal blocks are identical which allows the factorization of just one of them to be reused as to solve Equation (\ref{eq:k_el}) with three independent applications of forward/backward substitution for each  component of $\delta \aggvec{x}$.

\subsection{Collisions}

In the spirit of prior work \cite{teschner2005collision,Mcadams2011,mitchell2015non}, we process collisions by sprinkling a number of points on the surface of our volumetric model that we refer to as \textit{collision proxies}. These collision proxies can either be selected among the surface vertices of a conforming volumetric mesh, or simply embedded in the mesh in the sense that each of their locations is barycentrically interpolated from nodes of the containing element. That is, we can represent the location of the $j^{\text{th}}$ proxy point, $\vect{p}_j, \; j=1,\ldots,m$, as a weighted sum of all $n$ mesh vertex locations
$$
\vect{p}_j = \sum_{i=1}^n w_{(j,i)} \vect{x}_i \;\text{ where }\; \sum_{i=1}^n w_{(j,i)} = 1
$$
Note that this vector can be decomposed into a corresponding equation for each component, $v=1,\ldots,3$:
\begin{equation} \label{eq:p_j_v}
\vect{p}_j\indx{v} = \sum_{i=1}^n w_{(j,i)} \vect{x}_i\indx{v} \;\text{ where }\; \sum_{i=1}^n w_{(j,i)} = 1
\end{equation}
where $w_{(j,i)}$ is the weight of the $i^{\text{th}}$ vertex for the $j^{\text{th}}$ proxy. We expect that only 4 components of $w_{(j,i)}$ for any given $j$ are non-zero, corresponding to the vertices of the tetrahedron containing the proxy.
At each time step of the simulation, we will make a check to determine if any of these proxies are located in a prohibited region (e.g., inside a kinematic colliding object). Supposing that proxy $\vect{p}_j$ is inside a prohibited region, we will use the geometric representation of the obstacle (typically an implicit surface), to project the proxy location to the colliding region's surface. We label that point on the obstacle surface $\vect{t}_j$. We then instantiate a short lived, zero-restlength spring connecting $\vect{p}_j$ and $\vect{t}_j$. These springs will contribute to the energy of the system that we seek to minimize. We can write the energy contribution due to collisions concretely as
$$
E_{\text{col}} = \sum_{j=1}^m \frac{c_j}{2}||\vect{p}_j - \vect{t}_j||_2^2 \cdot \delta_j
$$
where $c_j$ is the stiffness coefficient of the $j^{\text{th}}$ proxy and $\delta_j$ is the indicator function
\begin{equation} \label{eq:indic}
\delta_j = \begin{cases} 
      0 & j^{\text{th}} \text{ proxy is not in collision} \\
      1 & j^{\text{th}} \text{ proxy is in collision}
   \end{cases}
\end{equation}
We can further decompose this by components:
\begin{equation} \label{eq:e_col_j}
E_{\text{col}} = \sum_{v=1}^3\sum_{j=1}^m \frac{c_j}{2}\left(\vect{p}_j\indx{v} - \vect{t}_j\indx{v}\right)^2 \cdot \delta_j
\end{equation}
We can then substitute (\ref{eq:p_j_v}) into (\ref{eq:e_col_j}) and achieve:
\begin{equation} \label{eq:E_col_C}
E_{\text{col}} = \sum_{v=1}^3\frac{1}{2}\left(\mat{W}\aggvec{x}\indx{v} - \aggsca{t}\indx{v}(\aggvec{x})\right)\tran C(\aggvec{x}) \left(\mat{W}\aggvec{x}\indx{v} - \aggsca{t}\indx{v}(\aggvec{x})\right)
\end{equation}
where the diagonal matrix $ \mat{C}(\aggvec{x})$ satisfies $[\mat{C}(\aggvec{x})]_{jj} = c_j \cdot \delta_j$ and $\mat{W}_{ji} = w_{(j,i)}$. Let us highlight two subtle but important points about equation (\ref{eq:E_col_C}): First, the only components of the equation that are dependent on $\aggvec{x}$ are $\mathbf{t}\indx{v}(\aggvec{x})$ and $C(\aggvec{x})$, with the dependence of the latter being due to proxies being flagged as active or inactive as a function of their placement. Second, the three components ($x$, $y$, and $z$) are separable and independent, just as we saw with the global step of projective dynamics. These observations suggest we follow the path of Projective Dynamics derivation further. We can write $E_{\text{col}}$ as an energy-minimization problem:
$$
E_{\text{col}} = \min_{\aggvec{t} :  \text{collision-free}} \sum_{v=1}^3\frac{1}{2}\left(W\aggvec{x}\indx{v} - \aggvec{t}\indx{v}\right)\tran \mat{C} \left(W\aggvec{x}\indx{v} - \aggvec{t}\indx{v}\right)
$$
where the ``projections'' $\vect{t}_j$ are selected among all \emph{collision-free} locations in the ambient space, as to minimize this energy. Conceptually, this suggests that by freezing $\aggvec{t}$ and $\mat{C}$ to specific values (determined by collision detection) as part of the local step, we retain both the stability traits of Projective Dynamics, and the property that this expression becomes a quadratic function. We denote this by $\hat{E}_{\text{col}}(\aggvec{x})$ and this energy term can be folded into the Newton scheme in Equation (\ref{eq:nr}) in the global step.

\section {Proposed Method}

We present our method by first partitioning our mesh into a collision-prone and a collision-safe region. We then use a Schur complement method to craft a numerical solution that concentrates on the collision prone region. Finally, we present a nested iteration that can refine the solution in the vicinity of collisions, at low cost.


\subsection{Broader context}

The power of projective dynamics is largely due to the ability to pre-factorize the system stiffness matrix using a Cholesky Factorization. Once this matrix is factorized, performing the global step of Projective Dynamics only incurs the cost of a single forward and backward substitution. However, as we discussed, when the simulation involves collisions, the system matrix changes at each step, compromising the ability to use a constant, pre-factored matrix.
Let us start by taking a closer look at the total energy equation of the global step, which is the sum of the energy due to elastic deformation and the energy due to collisions:
\begin{equation} \label{eq:e_tot}
\hat{E}_{\text{tot}} = \hat{E}_{\text{el}} + \hat{E}_{\text{col}}
\end{equation}
Differentiating (\ref{eq:e_tot}) once, we see the total forces (the complete right hand side of (\ref{eq:k_el}) :
$$
-\frac{\partial E_{\text{tot}}}{\partial \vect{x}} = \aggvec{f}_{\text{tot}} = \aggvec{f}_\text{el}(\aggvec{x}) - \mat{W}\tran \mat{C}(\aggvec{x})(\mat{W} \aggvec{x} + \aggvec{t})
$$
and differentiating again we see the complete left hand side of (\ref{eq:k_el}):
$$
\frac{\partial^2E_{\text{tot}}}{\partial \vect{x}^2} = \mat{K} + \mat{W}\tran \mat{C}(\aggvec{x}) \mat{W}
$$
Unfortunately, it is the case that our constant matrix used in the global step has been polluted by terms that depend on $\aggvec{x}$. There are two straightforward options we can consider: One option is to  perform a refactorization of the matrix at each timestep. Given that the matrices we are targeting will have in the order of $10^5$ nodal degrees of freedom, we cannot tolerate the pre-factorization cost at each time step for an interactive application. A second option is to use an iterative approach to solve the system without factorization. In section \ref{sec:results} we discuss when this is appropriate, but also note that convergence may then suffer for high resolution models.
Perhaps another option would be to observe that the matrix $\mat{W}\tran \mat{C}(\aggvec{x}) \mat{W}$ that depends on $\aggvec{x}$ is low rank. The rank of this collision matrix is a function of the number of active collisions at any one time. We could contemplate using methods for low-rank updates on factorized matrices. The problem with this is that even our ``low rank'' matrix has a rank in the hundreds to low-thousands for reasonable simulation scenarios. This would quickly yield an untenable proposition trying to manage such an effort with any low-rank update algorithm.

\begin{figure}
	\includegraphics[width=\columnwidth]{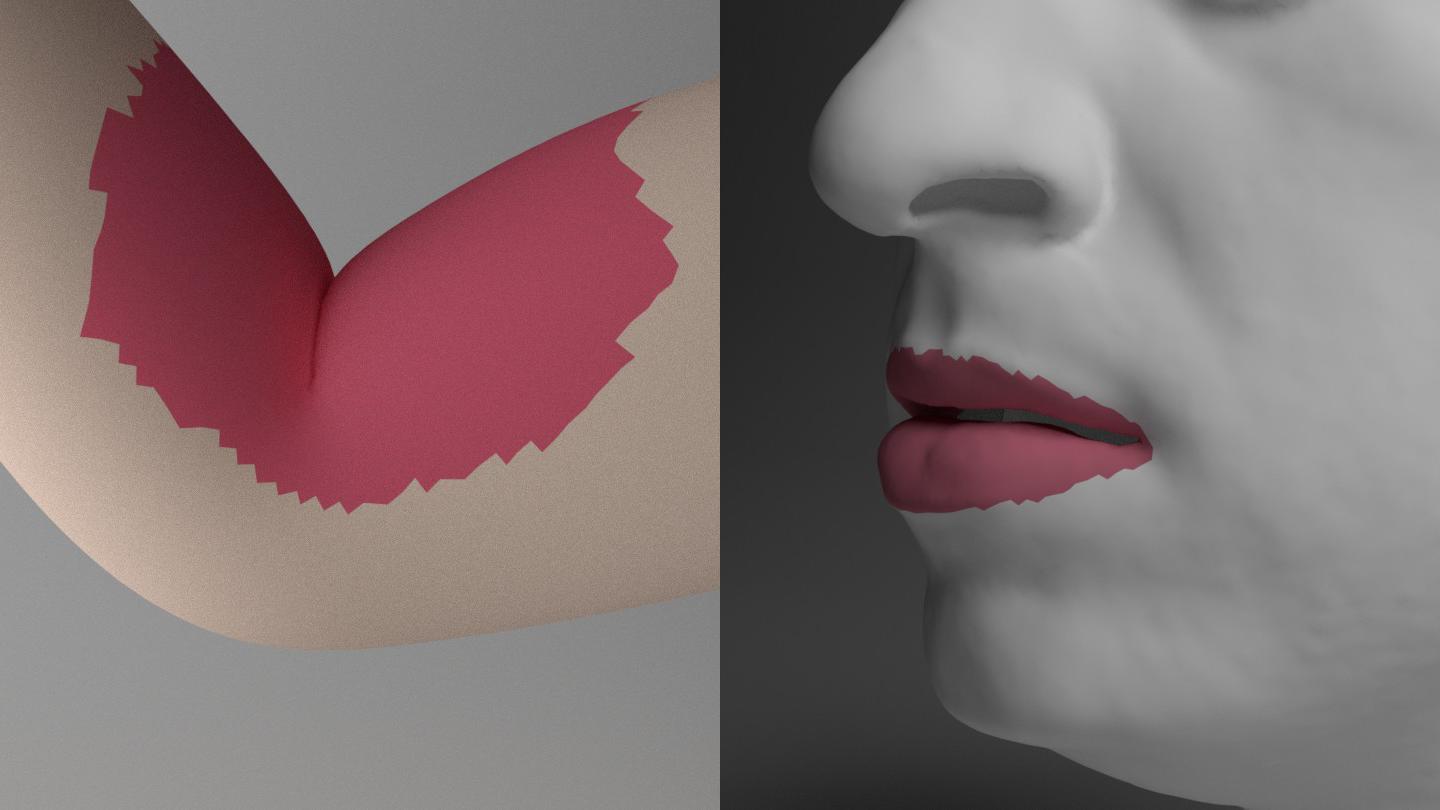}
    \caption{Our method requires an a-priori designation of a fraction of nodes as ``collision-prone''. For the models used in our examples, those regions are highlighted in red. For efficiency we aim to limit such nodes to a small subset (e.g. 5\%) of the total mesh vertices.}
    \label{fig:coll_regions}
\end{figure}

\subsection{Domain partitioning for the Global Step}

Motivated by these observations, we craft an approach that leverages our modeling hypotheses, namely:
\begin{enumerate}
    \item The fraction of the mesh prone to collision is a small subset ($<5\%$) of the simulated model, and
    \item The region where collisions may occur can be known a-priori.
\end{enumerate}

Consider the rank of the components of the global step matrix:
$$
\overbrace{\mat{K} + \underbrace{\mat{W}\tran \mat{C} \mat{W}}_{\text{rank }m}}^{\text{rank }n}
$$
Here, $n$ is the number of simulation mesh vertices. The part of the matrix that is changing, however, is a much smaller $m\times m$ submatrix. An upper bound on $m$ will be the cardinality of the union of vertices of tetrahedral elements that contain a collision proxy. As a practical matter, in our simulation of the face, $m$ is the number of degrees of freedom of the tetrahedral elements surrounding the lips, where in the arm model the same area is localized around the inner fold of the elbow joint, as shown in figure \ref{fig:coll_regions}. In the experimental examples presented in section \ref{sec:results}, models typical have in the order of 500K tetrahedra, 100K vertices, of which 1000-3000 vertices fit the criteria of anchoring a tetrahedral element that contains a collision proxy. 

\begin{figure}
\begin{center}
	\includegraphics[width=0.8\columnwidth]{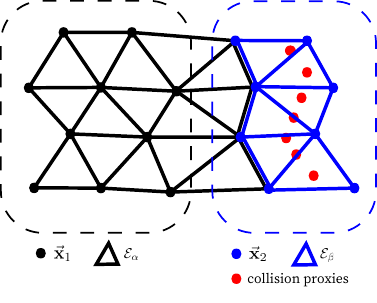}
	\caption{Simulation mesh partitioning. Collision proxies  shown in red. Elements that embed collision proxies form the set $\mathcal{E}_\beta$ (blue), while their complement is the collision-safe region $\mathcal{E}_\alpha$ (black). The collision-prone nodes $\aggvec{x}_2$ (blue) are those that appear in elements of $\mathcal{E}_\beta$, while all other (collision-safe) nodes are grouped in $\aggvec{x}_1$ (black).
	}
	\label{fig:x1_x2}
\end{center}
\end{figure}

Referring to figure \ref{fig:x1_x2} as an example, we can partition the full set of vertices, $\aggvec{x}$, into two subsets: 
$$
\aggvec{x} = \begin{pmatrix} \aggvec{x}_1 \\ \aggvec{x}_2 \end{pmatrix}
$$
where $\aggvec{x}_1$ contains the nodes \textit{not} in the immediate vicinity of collision proxies, and $\aggvec{x}_2$ contains the nodes that \textit{are} in the immediate vicinity of collisions. To further clarify, in figure \ref{fig:x1_x2}, $n$ is the total number of all vertices ($\aggvec{x}_1 \cup \aggvec{x}_2$), and $m$ is the number of vertices in $\aggvec{x}_2$.
Then, we  re-write the global step equation as follows
$$
    (\mat{K} + \mat{W}\tran \mat{C} \mat{W}) \delta \aggvec{x} = \aggvec{f}_{\text{tot}}
$$
in block form:
\begin{equation} \label{eq:block1}
    \begin{pmatrix*}[l] \mat{K}_{11} & \mat{K}_{12} \\ \mat{K}_{21} & \mat{K}_{22} + \mat{C}_{22}(\aggvec{x})\end{pmatrix*} 
    \begin{pmatrix} \delta \aggvec{x}_1 \\ \delta \aggvec{x}_2 \end{pmatrix} =
    \begin{pmatrix*}[l] \aggvec{f}_1 \\ \aggvec{f}_2 + d_2(\aggvec{x}) \end{pmatrix*}
\end{equation}
where
$
\mat{C}_{22}(\aggvec{x}) = \mat{W}\tran \mat{C}(\aggvec{x}) \mat{W}
$
and $d_2$ are the force components induced by collisions.
Based on the relative sizes of $\aggvec{x}_1$ and $\aggvec{x}_2$, we point out that the entire matrix is largely unchanged and remains constant, with only a very small subset of the matrix being dependent on the current vertex locations. Next, we capitalize on this structure and relative sizes by using a Schur complement method.

\subsection{Partial Cholesky Schur Complement Factorization}

Consider a linear system, $Ax=b$, where $A$ is symmetric. Partition $A$, $x$, and $b$  such that the system can be written in block format:
\begin{equation}\label{eq:axb_par}
    \begin{pmatrix} \mat{A}_{11} & \mat{A}_{12} \\ \mat{A}_{21} & \mat{A}_{22}\end{pmatrix} 
    \begin{pmatrix} x_1 \\ x_2 \end{pmatrix} =
    \begin{pmatrix} b_1 \\ b_2 \end{pmatrix}
\end{equation}
Suppose that $\mat{A}_{11}$ has the Cholesky factorization $\mat{A}_{11} = \mat{L}_1 \mat{L}_1\tran$. Careful multiplication will verify the following factorization of $A$:
\begin{equation}\label{eq:par_chol}
    \begin{pmatrix} \mat{A}_{11} & \mat{A}_{12} \\ \mat{A}_{21} & \mat{A}_{22}\end{pmatrix} = 
    \begin{pmatrix} \mat{L}_1 & 0 \\ \mat{A}_{21}\mat{L}_1\invtran & \mat{I} \end{pmatrix}
    \begin{pmatrix} \mat{I} & 0 \\ 0 & \Sigma \end{pmatrix}
    \begin{pmatrix} \mat{L}_1\tran & \mat{L}_1\invse \mat{A}_{12} \\ 0 & \mat{I} \end{pmatrix}
\end{equation}
where
$
\Sigma = \mat{A}_{22} - \mat{A}_{21} \mat{A}_{11}\invse \mat{A}_{12}
$
is known as the Schur complement. It is important to realize that the factorization in equation (\ref{eq:par_chol}) is nothing more than a partial Cholesky factorization, and is typically an intermediate step in the full Cholesky factorization of. The Intel\textsuperscript{\textregistered} MKL PARDISO library, which we use, can be invoked as to compute exactly such a factorization, providing the user with the express value of the Schur complement while retaining the partial triangular factors in its internal representation.

We should note that our partitioning of nodes into the subsets $\aggvec{x}_1$ and $\aggvec{x}_2$ does incur some slight sub-optimality relative to the stock (non-Schur) Cholesky factorization,
by constraining the degrees of freedom in $\aggvec{x}_1$  to appear strictly before those in $\aggvec{x}_2$. Our experiments however indicate that any deterioration in sparsity of the resulting factors was less than $10\%$ in all of our examples.

\subsection{Towards an Accelerated Solution}


The Intel\textsuperscript{\textregistered} MKL PARDISO sparse factorization library provides a highly optimized CPU-based implementation of the partial factorization shown in equation (\ref{eq:par_chol}) and discussed above. The user provides as input the symmetric system matrix and the degrees of freedom that are to be maintained after the factorization. The library provides as output the (dense) Schur complement $\Sigma$ matrix, and maintains the partial triangular factors in internal representation. 

Now our system shown in (\ref{eq:axb_par}), after factorization, becomes:
\begin{equation}\label{eq:par_chol2}
    \underbrace{\begin{pmatrix} \mat{L}_1 & 0 \\ \mat{A}_{21}\mat{L}_1\invtran & \mat{I} \end{pmatrix}}_{\text{lower-tri}}
    \begin{pmatrix} I & 0 \\ 0 & \Sigma \end{pmatrix}
    \underbrace{\begin{pmatrix} \mat{L}_1\tran & \mat{L}_1\invse \mat{A}_{12} \\ 0 & \mat{I} \end{pmatrix}}_{\text{upper-tri}}
    \begin{pmatrix} \aggvec{x}_1 \\ \aggvec{x}_2 \end{pmatrix} =
    \begin{pmatrix} \aggvec{b}_1 \\ \aggvec{b}_2 \end{pmatrix}
\end{equation}

We can solve this in a series of steps:

\textbf{Step 1:} Solve the lower triangular system:
\begin{equation}
    \begin{pmatrix} \mat{L}_1 & \mat{0} \\ \mat{A}_{21}\mat{L}_1\invtran & \mat{I} \end{pmatrix}
    \begin{pmatrix} \aggvec{y}_1 \\ \aggvec{y}_2 \end{pmatrix} =
    \begin{pmatrix} \aggvec{b}_1 \\ \aggvec{b}_2 \end{pmatrix}
\end{equation}
Being a lower triangular matrix, this can be solved quickly by forward substitution on the CPU using Intel\textsuperscript{\textregistered} MKL PARDISO optimized forward substitution algorithm (PARDISO ``phase 331'').

\textbf{Step 2:} Solve the system:
\begin{equation}
    \begin{pmatrix} \mat{I} & \mat{0} \\ \mat{0} & \Sigma \end{pmatrix}
    \begin{pmatrix} \aggvec{z}_1 \\ \aggvec{z}_2 \end{pmatrix} =
    \begin{pmatrix} \aggvec{y}_1 \\ \aggvec{y}_2 \end{pmatrix}
\end{equation}
It is trivial to see that $\aggvec{z}_1 = \aggvec{y}_1$ and that $\aggvec{z}_2 = \Sigma\invse \aggvec{y}_2$. We will momentarily defer discussion of how to solve this until completing the description of the full solution.

\textbf{Step 3:} Finally, Solve the system:
\begin{equation}
    \begin{pmatrix} \mat{L}_1\tran & \mat{L}_1\invse \mat{A}_{12} \\ \mat{0} & \mat{I} \end{pmatrix}
    \begin{pmatrix} \aggvec{x}_1 \\ \aggvec{x}_2 \end{pmatrix} =
    \begin{pmatrix} \aggvec{z}_1 \\ \aggvec{z}_2 \end{pmatrix}
\end{equation}
Being an upper triangular matrix, this can be solved quickly by backward substitution on the CPU using MKL PARDISO optimized backward substitution algorithm (``phase 333'').

 
 
 
 At first glance, a concern might be that since the overall matrix in our application is changing with each time step due to collisions, that we may still have to recompute this partial factorization at each timestep to produce a new $\Sigma$. Fortunately this is not the case.  To see this, refer to (\ref{eq:block1}) and consider the Schur complement of the $K$ matrix \textit{without} any collision forces. \textit{Without} collision forces, the Schur complement of the block $K$ matrix would be 
\begin{equation}\label{eq:k22}
\Sigma_{\text{no-col}} = \mat{K}_{22} - \mat{K}_{21}\mat{K}_{11}\invse \mat{K}_{12}
\end{equation}
In the presence of collisions, the term $\mat{K}_{22}$ has been replaced by $\mat{K}_{22} + \mat{C}_{22}(\aggvec{x})$. Because $\mat{K}_{22}$ only appears on the right hand side of (\ref{eq:k22}) as a lone term, we can see that in the presence of collisions, we can simply add the $\mat{C}_{22}(\aggvec{x})$ term to yield:
\begin{align*}
\Sigma_{\text{col}} &= \mat{K}_{22} - \mat{K}_{21}\mat{K}_{11}\invse \mat{K}_{12} + \mat{C}_{22}(\aggvec{x}) \\
                   &= \Sigma_{\text{no-col}} + \mat{C}_{22}(\aggvec{x})
\end{align*}
This means that we can compute the partial factorization only once in the absence of collisions and retain a $\Sigma_{\text{no-col}}$ matrix, to which we can add $\mat{C}_{22}(\aggvec{x})$ at each time step. This addition is performed as a purely \emph{additive}  update to the Schur Complement, and is very efficient.
Now knowing that it is easy to produce the correct $\Sigma$ matrix at each timestep, we return to describing an efficient solution to $\Sigma \aggvec{z}_2 = \aggvec{y}_2$. Recalling that the expected size of $\Sigma$ is approximately 1000-3000 degrees of freedom, we are presented with a slightly surprising opportunity that one might otherwise overlook. Although for the \emph{global, sparse matrix} it is not practical to repeat a factorization every time its entries change, for the \emph{local, dense} matrix $\Sigma$ the refactorization is a perfectly realistic and efficient option.


To support this point, let us explore the cost of factorizing $\Sigma$ and directly solving $\Sigma \aggvec{z}_2 = \aggvec{y}_2$. For purposes of illustration, we will consider a typical $\Sigma$ in our simulation having $m \approx 2000$ degrees of freedom. (dimension $\approx2000 \times 2000$). A full Cholesky factorization requires $\frac{1}{6}m^3$ floating point operations (FLOPS), or in our case will be $\approx\frac{8}{3}$ billion floating point operations (GFLOPS). Modern workstation-grade CPUs are capable of $2$+ trillion floating point operations (TFLOPS) per second, and modern high end GPUs are capable of approximately $12$ TFLOPS. This suggests we have the ability to factorize such a system in a budget of low number of milliseconds; such opportunity is not afforded to \emph{sparse} matrices, as their processing is often bound by memory bandwidth. For such sizes of dense matrices however, the cubic computational complexity is well counterbalanced by the (typical 100:1) ratio of possible arithmetic computations per memory access on  CPUs or GPUs.


With this ``order of magnitude'' calculus suggesting that we have a promising approach to achieving the desired performance, we describe our  implementation of Step 2. We solve this system on the GPU. At the start of simulation, we move the $\Sigma_{\text{no-col}}$ matrix to the GPU memory. At each timestep, we move the $\mat{C}(\aggvec{x})$ matrix and the $\aggvec{y}_2$ vector to the GPU. In case of self-collisions, we may need to also transmit to the GPU the embedding weight matrix $W$. Recall that $\mat{C}(\aggvec{x})$ is a diagonal matrix, so the bandwidth per timestep is small, and similarly $W$ is sparse. We then use highly efficient stock algorithms available in NVIDIA \textsf{cuSPARSE} and \textsf{cuSOLVER} libraries to a) perform the rank-$k$ update on the pre-calculated Schur complement matrix (using \textsf{cusparseScsrgemm2}), b) refactorize it on the fly (with \textsf{cusolverDnSpotrf}), and c) perform the forward and backward substitution to provide the solution (\textsf{cusolverDnSpotrs}), $\aggvec{z}_2$, which we stream back to main memory and proceed with Step 3 to complete the solution steps for the current time step.

\IncMargin{1em}
\begin{algorithm}[t]
\SetKwFunction{LocalStep}{LocalStep}\SetKwFunction{ComputeForces}{ComputeForces}
\SetKwFunction{DetectCollisions}{DetectCollisions}
\SetKwFunction{LocalStep}{LocalStep}
\SetKwInOut{Input}{input}\SetKwInOut{Output}{output}
\textbf{preliminary:} Schur-factorize:
\begin{equation*}
\begin{pmatrix} \mat{A}_{11} & \mat{A}_{12} \\ \mat{A}_{21} & \mat{A}_{22}^{(\alpha)}\end{pmatrix} = 
    \begin{pmatrix} \mat{L}_1 & 0 \\ \mat{A}_{21}\mat{L}_1\invtran & \mat{I} \end{pmatrix}
    \begin{pmatrix} \mat{I} & 0 \\ 0 & \Sigma_\alpha \end{pmatrix}
    \begin{pmatrix} \mat{L}_1\tran & \mat{L}_1\invse \mat{A}_{12} \\ 0 & \mat{I} \end{pmatrix}
\end{equation*}

\Input{partially factorized stiffness matrix with Schur Complement; updated Dirichlet node positions}
\Output{Correction amount $\aggvec{u} = \begin{pmatrix}\aggvec{u}_1 \\ \aggvec{u}_2 \end{pmatrix}$}
\BlankLine
\For{$i\leftarrow 1$ \KwTo $outer\_iters$}{
1. $\aggmat{R}_\alpha \leftarrow $ \LocalStep() \\
2. $\begin{pmatrix*}[l] \aggvec{f}_1 \\ \aggvec{f}_2^{(\alpha)}\end{pmatrix*} \leftarrow$ \ComputeForces() \tcp*[h]{only from $\hat{E}_\alpha$} \\
3. Solve $\begin{pmatrix} \mat{L}_1 & \mat{0} \\ \mat{A}_{21}\mat{L}_1\invtran & \mat{I} \end{pmatrix}
    \begin{pmatrix} \aggvec{y}_1 \\ \aggvec{y}_2 \end{pmatrix} =
    \begin{pmatrix*}[l] \aggvec{f}_1 \\ \aggvec{f}_2^{(\alpha)}\end{pmatrix*}$ via ForwardSub \\
3.1. We expect: $\begin{pmatrix} \aggvec{y}_1 \\ \aggvec{y}_2 \end{pmatrix} =
\begin{pmatrix*}[l] \mat{L}_1\invse \aggvec{f}_1 \\ \aggvec{f}_2^{(\alpha)} - \mat{A}_{21}\mat{A}_{11}\invse \aggvec{f}_1 \end{pmatrix*}$ \\
3.2 $\mathbf{\tilde{f}}_2^{(\alpha)} = \aggvec{f}_2^{(\alpha)} - \mat{A}_{21}\mat{A}_{11}\invse \aggvec{f}_1 = \aggvec{y}_2$

4. \For{$j\leftarrow 1$ \KwTo $inner\_iters$}{

4.1 $\mat{C}(\aggvec{x_2}) \leftarrow$ \DetectCollisions$(\aggvec{x}_2)$  \\
4.2. $ \aggmat{R}_\beta \leftarrow $ \LocalStep() \\
4.3 $\mat{H} \leftarrow  \Sigma_\alpha + \mat{A}_{22}^{(\beta)} + A_{22}^{(\text{col})}$ \\
4.4 $\aggvec{g} \leftarrow \mathbf{\tilde{f}}_2^{(\alpha)} + \aggvec{f}_2^{(\beta)}(\aggvec{x}_2, \aggmat{R}_{\beta}) + \aggvec{f}_2^{(\text{col})}(\aggvec{x}_2)$ \\
4.5 Solve $\mat{H} \aggvec{u}_2 = \aggvec{g}$ \\
4.6 $\aggvec{x}_2 += \aggvec{u}_2$ \\
4.7 $\mathbf{\tilde{f}}_2^{(\alpha)} -= \Sigma_\alpha \aggvec{u}_2$
}
5. Solve $L_1\tran \aggvec{u}_1 = \aggvec{y}_1$ using BackwardSub 
}
\caption{Optimized solve with collisions}\label{alg:inneriter}
\end{algorithm}\DecMargin{1em}
\subsection{Further Optimization of the Solution}

A frequent observation in simulation of volumetric objects with collisions is that the non-linear, highly volatile penalty terms are the main contributor to the need for a large number of iterations at each time step to achieve convergence. In particular, it is the changing nature of collision proxies alternating between being active and inactive during iteration. Although all of these volatile behaviors are localized, traditionally the cost that we pay is global.

In this section, we take the opportunity to investigate the possibility to completely restrict the iterative computation so that it only takes place in the immediate vicinity of the collision prone region. To begin, refer again to figure \ref{fig:x1_x2}, observing the two partitions: 
\begin{enumerate}
\item The elements are partitioned into black elements, $\elemset{E}_\alpha$, which are elements that do not contain collision proxies, and blue elements, $\elemset{E}_\beta$, which are elements that \textit{do} contain collision proxies. 
\item Vertices of the collision-prone elements $\elemset{E}_\alpha$ will be labeled the collision-prone set $\aggvec{x}_2$ (in blue); their complement will be the collision-safe nodes $\aggvec{x}_1$ (in black).
\end{enumerate}


In this partitioning the element set $\elemset{E}_\alpha$ is associated with vertices from \textit{both} $\aggvec{x}_1$ and $\aggvec{x}_2$, while $\elemset{E}_\beta$ is associated with vertices \textit{only} from $\aggvec{x}_2$.
We also refer back to equation (\ref{eq:e_tot}) that defines the total energy, recalling that the quasistatic solution can be written as a minimization problem under the context of Projective Dynamics (as in (\ref{eq:min_e})):
\begin{align}
\min_{\aggvec{x}}E(\aggvec{x}) &=\min_{\aggvec{x}} \left(E_{\text{el}}(\aggvec{x}) + E_{\text{col}}(\aggvec{x})\right) \\
              &\hspace*{-.4in}= \min_{\aggvec{x}, \aggmat{R}} \left(\hat{E}_{\text{el}}(\aggvec{x}, \aggmat{R}) + E_{\text{col}}(\aggvec{x_2})\right) \\
              &\hspace*{-.4in}= \min_{\aggvec{x}_1, \aggvec{x}_2, \aggmat{R}_\alpha, \aggmat{R}_\beta} \left(\hat{E}_{\alpha}(\aggvec{x}_1, \aggvec{x}_2, \aggmat{R}_{\alpha}) + \hat{E}_{\beta}(\aggvec{x}_2, \aggmat{R}_{\beta}) + E_{\text{col}}(\aggvec{x}_2)\label{eq:sep_min}\right)
\end{align}
The final line above separates the equation into contributions from $\elemset{E}_\alpha$ and $\elemset{E}_\beta$, being careful to note that the first ($\alpha$) term is a function of $\aggvec{x}_1$ and $\aggvec{x}_2$ while the second ($\beta$) term is a function only of $\aggvec{x_2}$.

To see how we can solve the global step in a nested iteration, first assume that the $\aggmat{R}_\alpha$ and $\aggmat{R}_\beta$ have been fixed by the local step. We can then rewrite (\ref{eq:sep_min}) with fixed rotations as a nested minimization:
\begin{equation}
E(\aggvec{x}) = \min_{\aggvec{x}_2} \Big\{ \underbrace{\min_{\aggvec{x}_1} \hat{E}_{\alpha}(\aggvec{x}_1, \aggvec{x}_2, \aggmat{R}_{\alpha})}_{\tilde{E}_\alpha(\aggvec{x}_2, \aggmat{R}_\alpha)} + \hat{E}_{\beta}(\aggvec{x}_2, \aggmat{R}_{\beta}) + E_{\text{col}}(\aggvec{x}_2) \Big\}
\end{equation}

Our method solves (\ref{eq:sep_min}) by a nested iteration that leverages the opportunity to do additional processing on the collision affected region ($\beta$) in an inner loop without compromising the the correctness of the solution in the non-collision affected region ($\alpha$). Our approach involves these steps:

\begin{description}
\item[Preamble of outer loop] We optimize only rotations $\aggmat{R}_\alpha$ in the collision-safe region, keeping all other variables fixed. 
\item[Inner loop] Treating  $\aggmat{R}_\alpha$ as fixed, we use the Schur Complement to express the minimum (over $\aggvec{x}_1$) of $\hat{E}_\alpha$ as a function, $\tilde{E}_\alpha$, of only $\aggvec{x}_2$ and $\aggmat{R}_\alpha$ (the latter being a constant). This is equivalent to the elimination of $\mathbf{x}_1$ from the global step via the Schur Complement, as described in the previous section. The resulting energy is only a function of $\aggvec{x}_2$ and $\aggmat{R}_\beta$ at this point, and we iterate on it in the style of Projective Dynamics -- freezing each of $\aggvec{x}_2$ and $\aggmat{R}_\beta$ and optimizing over the other -- combined with collision detection and update of proxies at the local step. 
\item[Conclusion of outer loop] We reconstruct the solution for the collision-safe region via backward substitution. 
\end{description}

\begin{figure*}[t]
\center{	\includegraphics[height=3in]{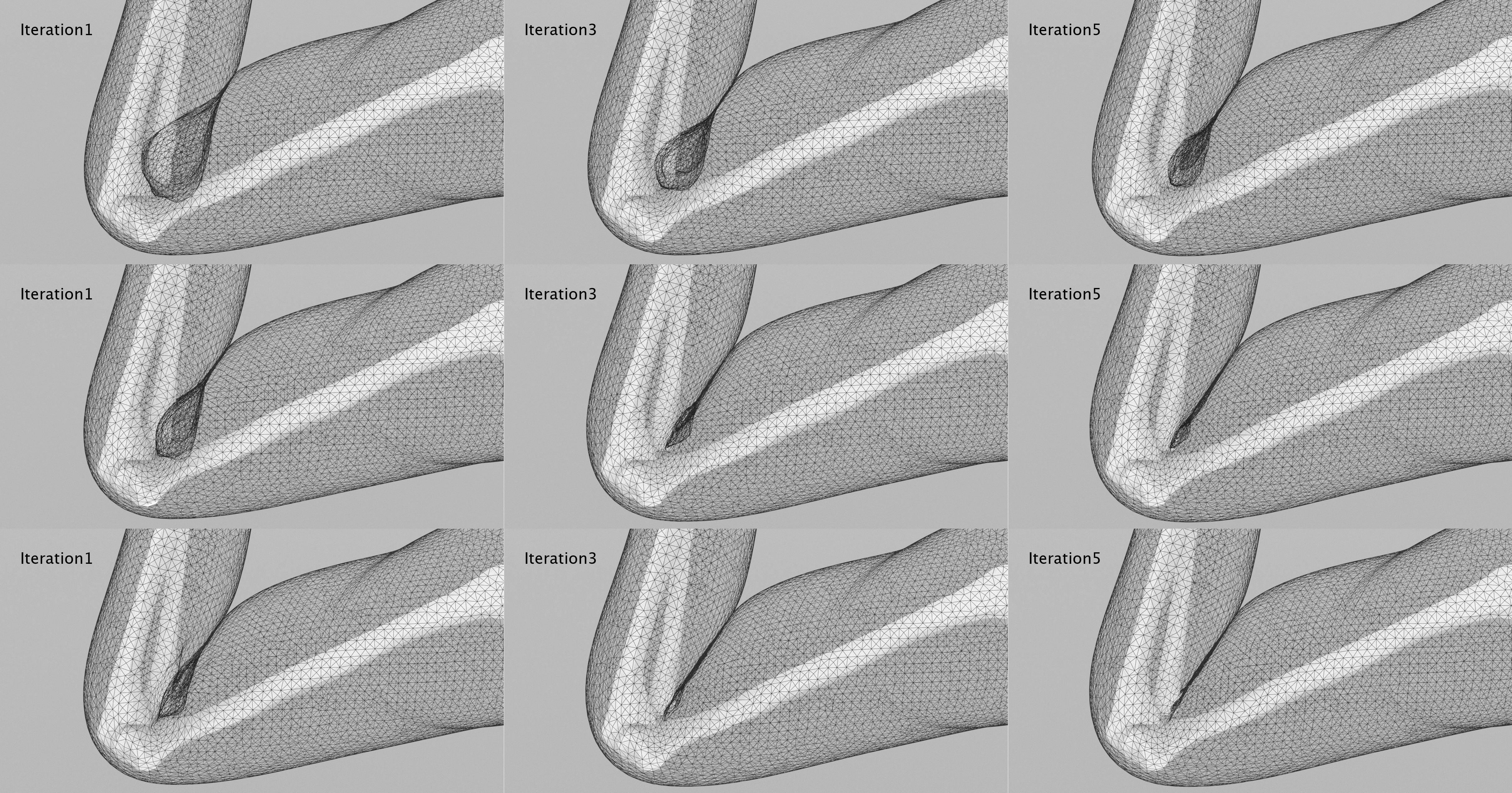}
}    \caption{The arm is bent with collisions disabled, and then collisions are turned on, forcing the mesh to untangle from the collided state. One inner iteration per PD loop is illustrated on the top, 3 inner iterations in the middle row, and 5 inner iterations per PD loop on the bottom.}
    \label{fig:inner_iters}
\end{figure*}

This nested iterative solution procedure is captured in Algorithm \ref{alg:inneriter}, where the specific utilization of the MKL PARDISO library is highlighted. From the inner loop, update of the Schur-derived Hessian matrix $H$, its dense factorization, and the solution for the local update $\aggvec{u}_2$, which are are hosted on the GPU.
As noted above, we perform steps 4.2 through 4.4 of the algorithm on the GPU. We point out that we make the choice of skipping the calculation of $\aggmat{R}_\beta$ in step 1 and instead doing it inside the inner loop before step 4.1. Updating $\aggmat{R}_\beta$ as part of each inner iteration will improve the rate of convergence, with the minimal extra cost of updating the small number of collision-affected rotations during each inner iteration. In our experience, this extra computation pays off in convergence rates as Figure \ref{fig:inner_iters} illustrates.


\section {Results and Evaluation}\label{sec:results}

\begin{figure}[b]
\begin{center}
	\includegraphics[width=\columnwidth]{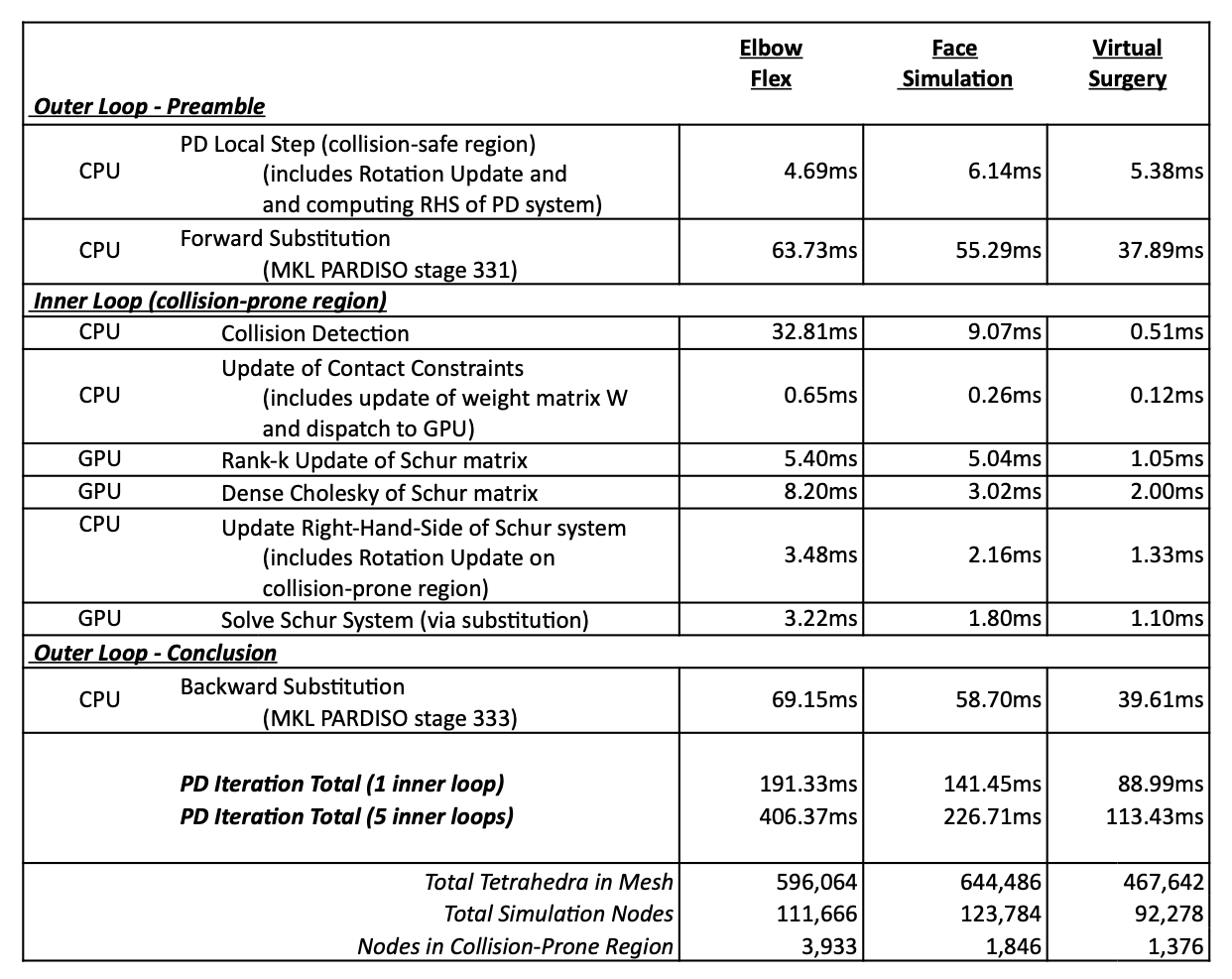}
	\vspace*{-.2in}
	\caption{Timing results of our three featured benchmarks.}
	\label{fig:benchmark}
\end{center}
\end{figure}

We  demonstrate  our method on three simulation scenarios, all of which achieve interactive performance with resolutions in the order of half million elements. We also perform a comparison of our direct solver to a GPU optimized iterative scheme \cite{Komaritzan2019}.



All our tests were on an Intel Core i9-9940X CPU @ 3.30GHz and a NVIDIA TITAN X GPU. See Figure \ref{fig:benchmark} for detailed benchmark results.
In our implementation, all outer loop calculations are run on the CPU including the Projective Dynamics local step, forward substitution and backward substitution. Inside the inner loop, we perform collision detection, update contact constraints, and update the right-hand side of the Schur system on the CPU. The other steps, including rank-$k$ update and dense Cholesky factorization are done on the GPU. For the local step on the CPU, we leverage AVX512 SIMD instructions to vectorize our code, including the update of best-fit rotations and computation of elemental forces.

\subsection{Test \#1: Arm Flexing at Elbow Joint}
The first example we examine involves simulating a human arm bending at the elbow joint. In this case, self collisions only occur on the flesh surface on the interior side of the joint. The arm model is embedded in a tetrahedron mesh with 644,486 elements and 111,666 simulation nodes. The potential colliding region is predetermined and marked with collision proxies, which results in 3,933 collision-prone nodes - about 3.5\% of the total simulation nodes, as seen in Figure \ref{fig:coll_regions} (left). Skeletal bones are attached to the simulation mesh by zero restlength springs uniformly sampled over the bone surface. The scripted motion bends the elbow joint by 2.5 degrees per animation frame, before coming to a stop at a pose that creates significant self-collision. Collision detection and response uses rest-pose levelset representations \cite{Mcadams2011} of the arm to detect interpenetration and instance collision springs

\begin{figure}[b]
    \includegraphics[width=\columnwidth]{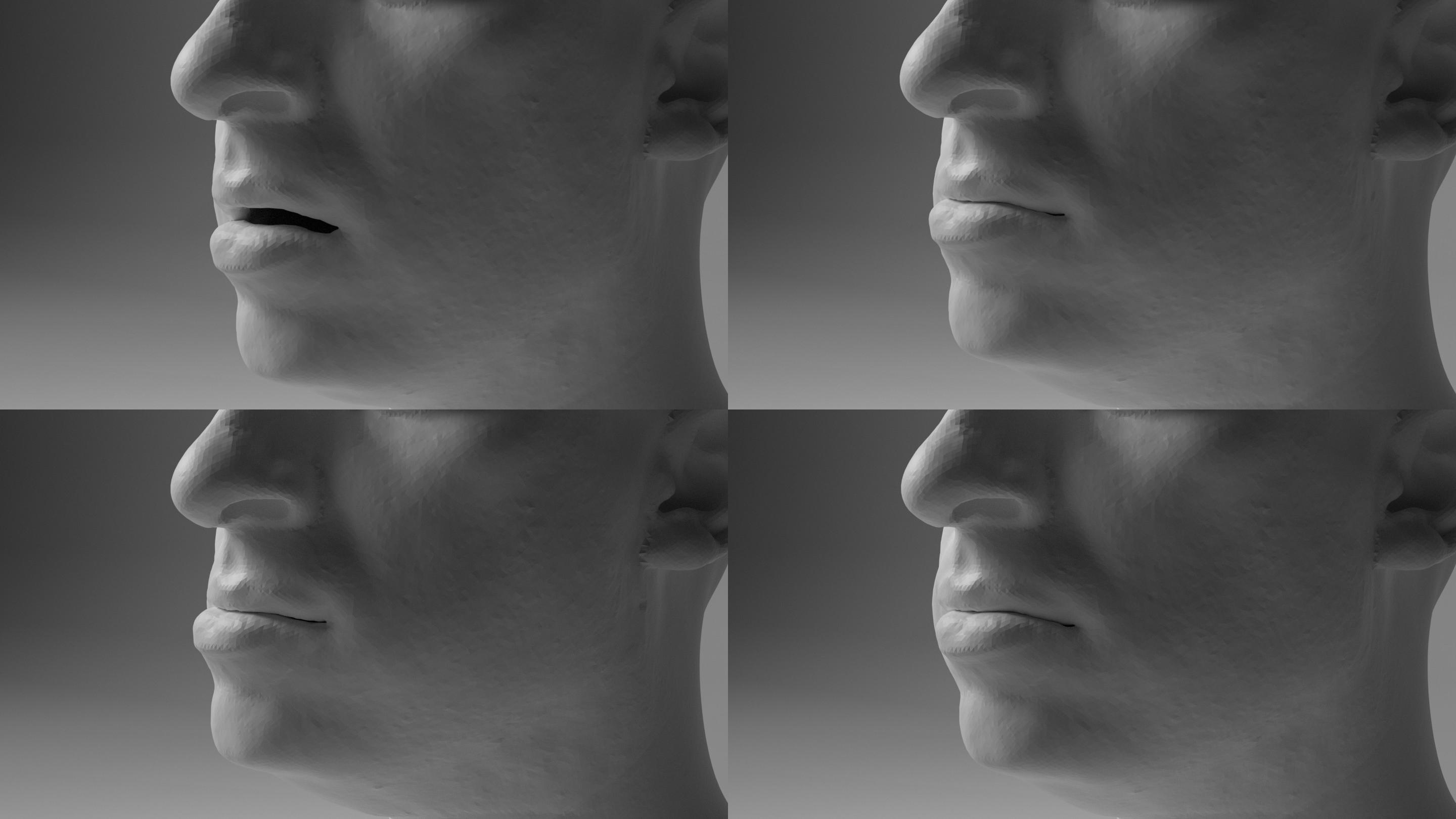}
    \caption{As the jaw moves, the lips engage in self-collision which is efficiently resolved. This face model contains 644k tetrahedral elements and 124k nodes, of which 1,846 are collision-prone. }
    \label{fig:face_manip}
\end{figure}
As can be seen in our supplemental video, using a direct solver for the global step allows our solver to produce a smooth and visually converged animation even with a single Projective Dynamics loop per animation frame. We experimented with both 1 inner-loop of localized update of element rotations in the collision-prone region per global PD iteration, and 3 or 5 inner-loops which only modestly adds to the solver cost (most of the added cost comes from the repeated detection step) but significantly improves the convergence in strenuous contact cases, as the test in Figure \ref{fig:inner_iters} where the elbow is brought to a sharp angle with collisions disabled, and attempts to disentangle from this state when collision response is again enabled. Even in challenging frames of this animation, we achieve 5fps with 1 inner loop, and 2.5fps with 5 inner loops. 

\subsection{Test \#2: Face Simulation with Self Collision at Lips}

The next case we look at, shown in figure \ref{fig:face_manip}, is a human face simulation, where self collisions occur purely around the lip region. With 644,486 embedding tetrahedral elements, there are 123,784 simulation nodes, only 1,846 of which are contained in the collision prone region. Here the ratio of collision-prone nodes is only 1.5\%. We achieve 5-7fps throughout this animation.

\subsection{Test \#3: Surgical Cleft Lip and Palate Simulation}
\begin{figure}[t]
\begin{center}
	\includegraphics[width=1.0\linewidth]{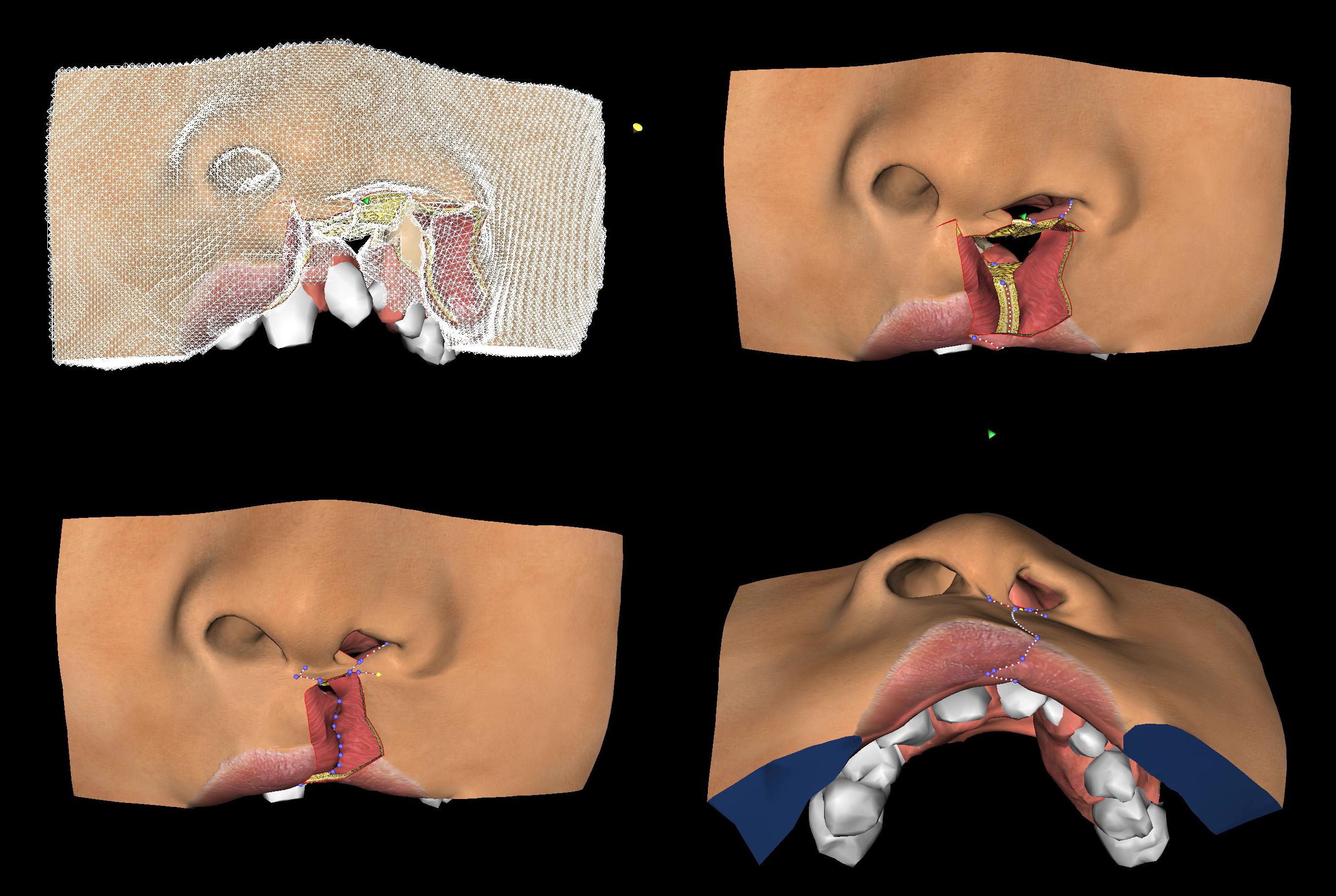}
\end{center}
	\caption{Frames from an interactive cleft lip surgery simulator.}
	\label{fig:surgery}
\end{figure}

Finally, we apply our algorithm in a virtual surgery simulator where a volumetric facial flesh mesh from a patient with cleft lip and palate is discretized into 503,910 tetrahedra and 97,249 simulation nodes. At some point during surgical manipulation, tissue is excised, leaving behind a simulation mesh with 467K tetrahedra and 92K vertices, as reported in Figure \ref{fig:benchmark}. 
In our test, only collisions between the deformable flesh and the rigid teeth and maxilla are processed, as the surgical repair being modeled relies on comprehensive suturing rather than self-collision to create the final repair and closure. With this hypothesis, we mark 1,434 collision prone nodes in the designated area inside of the lip. Screen captures from the interactive simulator  are shown in figure \ref{fig:surgery}.

To realistically simulate the elastic response of the simulated model, we note that human skin, and tissue beneath it, presents a unique bi-phasic behavior where a drastic increase of the stiffness of skin and tissue is observed when the stress is beyond a certain threshold, thus preventing further stretching of the tissue beyond a certain extent.
To take into account this effect of human skin, we modified the force density function $\Psi$ to the following form:
\begin{align*}
\Psi^\prime(\mat{F}) = \min_{\mat{Q}\in S} \mu^\prime || \mat{F} - \mat{Q} ||_F^2 +  \min_{\mat{R} \in \text{SO}(3)}  \mu || \mat{F} - \mat{R} ||_F^2\\
S = \{\mat{A}\in\mathbb{R}^{3\times3} \text{ s.t } \sigma_{\min}<\sigma_i(\mat{A})<\sigma_{\max}\}
\end{align*}
Where $\mat{F}$ is still the deformation gradient, $\sigma_i(\mat{A})$ is the $i$th singular value of $\mat{A}$, and $\sigma_{\min}, \sigma_{\max}$ are the lower and upper limits that we wish to allow our principal strain to assume. Scalar $\mu^\prime$ is the increased stiffness when tissue enters the bi-phasic regime. 
Similar to the PD formulation for corotated elasticity, we have
\begin{align*}
    E_e(\mat{F_e})^\prime &= \min_{\mat{R_e}\in\text{SO}(3), Q\in S} 
    \text{Vol}_e(\mu^\prime || \mat{F}_e - \mat{Q}_e ||_F^2 +   \mu || \mat{F}_e - \mat{R}_e ||_F^2)\\
    &=\min_{\mat{R_e}\in\text{SO}(3), Q\in S}\hat{E}_e^\prime(\mat{F}_e, \mat{R}_e, \mat{Q}_e)
\end{align*}
Again considering all $R_e$ and $Q_e$ momentarily as independent variables, the energy over all elements can be modified to :
$$
\hat{E}^\prime(\aggvec{x}, \aggmat{R}, \aggmat{Q}) = \sum_e \hat{E}^\prime_e(F_e(\aggvec{x}, R_e, Q_e))
$$
And the discrete energy is:
$
E^\prime(\aggvec{x}) = \min_{\aggmat{R}, \aggmat{Q}}\hat{E}^\prime(\aggvec{x},\aggmat{R},\aggmat{Q}).
$
This energy is fully compatible with the PD paradigm, and the minimization of $Q_e$ can be performed by computing SVD of $F_e$ and clamping its singular values to the bounds $\sigma_{\min} ,\sigma_{\max}$ in the local step.

\subsection{Comparison with GPU-optimized iterative solver \cite{Komaritzan2019}}

We performed a comparison with iterative solvers that could be natural alternatives to our method, especially for models of more modest resolution and detail. We focused on the GPU-accelerated PCG solver of the recent Fast Projective Skinning technique \cite{Komaritzan2019} as the most promising recent technique in terms of efficiency and features. Although the highest resolution model in their demos (91K tetrahedra) has 6-7 times fewer tets than our target meshes, they demonstrated real-time performance for models of that scale, making it possible that their technique might scale up to the half-million elements we accommodate. Although we did not have an end-to-end comparison due to differences in collision processing components and their use of embedding vs. conforming meshes in our approach, we performed a study of the comparative convergence efficiency of the two methods in the 500K element regime. Our findings are summarized below, and we have included a video with several comparative benchmarks; we should however clarify that if one is targeting resolutions below 100K tetrahedral elements, even though both methods would in principle yield interactive performance, the GPU-based PCG solver would not require a prescription of collision regions, and should be preferred due to its generality. 

We focused our comparison purely on the solver stage, excluding any runtime cost of assembling or updating the global step matrix or performing collision detection. We also modified their solver \cite{Komaritzan2019} to include support for embedded collision proxies, which are crucial to our examples. Jacobi preconditioning was used as suggested, although we did not experience any nontrivial acceleration, as our embedding meshes were perfectly regular. As seen in the supplemental video, we noticed that for high-resolution models the PCG solver required at least 100 (or more) iterations to start approaching the accuracy of the exact solver, and often exhibiting artifacts if inadequate convergence was reached in the global step. As an indication, the cost of 100 iterations (which was the bare minimum for acceptable or even stable convergence) in the  GPU PCG solver was 27ms for our \emph{elbow bending} example, and 21ms for our \emph{virtual surgery} scenario. These times are already 2-4x higher than our inner-loop costs (which produces exact solutions) in instances where multiple inner loops aid convergence, as the elbow simulation. Our method has to sustain the cost of a CPU forward/back-substitution at the beginning/end of each outer PD loop, which takes 70-130ms, but we have experimented with using stock GPU solvers for this stage which typically accelerate this stage by a factor of 3-4x. The direct solver offers the most accurate and robust convergence behavior, does not require parameter tuning to aid convergence (e.g. dynamics leads to much easier matrices for PCG than quasistatics), and is resilient to the stiffness of constraints such as bone attachments and collision penalty forces.




\section{Limitations \& Future Work}

We are naturally susceptible the restrictions of Projective Dynamics with respect to material models (primarily corotated elasticity) that can be accommodated. We did not demonstrate dynamic simulations, but such cases are straightforward extensions of our scheme. Our performance benefits are mostly realized when our target simulations are in the order of half-million elements, as in our demonstrations. Models of one order of magnitude smaller might be able to afford a full re-factorization in each PD step without losing interactivity, or enjoy adequate convergence with an iterative solver.
The most fundamental limitation of our work is our conscious assumption that contact regions are relatively small and static. Obviously there are many examples of highly relevant simulations that would not satisfy such preconditions. We look forward to investigating alternative methodologies for such problems, especially at even higher scale and resolution, such as Multigrid technqiues.


\printbibliography                

\end{document}